\documentclass[aps,prl,frontmatterverbose,reprint,amsmath,amssymb]{revtex4-2}
\usepackage{graphicx} % Required for inserting images
\usepackage{color}
\usepackage{amsmath}
\usepackage[abs]{overpic}

\begin{document}
\title{Excitation of subradiant states in atom chains by surface plasmon waves}
\author{Giorgio Vimercati \& Nicola Piovella}
\affiliation{Dipartimento di Fisica "Aldo Pontremoli", Universit\`{a} degli Studi di Milano, Via Celoria 16, I-20133 Milano, Italy}

\date{\today}

\begin{abstract}
Subradiant states of a finite subwavelength atomic chain in free space offer a powerful resource for photon storage. Yet their preparation and control remain challenging experimentally. A standard technique is to drive the atoms by a laser and observe the florescence light of the long-lived subradiant modes of excitation after the laser is switched off. However, these modes are highly degenerate and form an almost continuous band in the subradiant manifold. Here we propose a new method of exciting a given subradiant state by coupling atoms to a surface plasmon wave propagating along the interface with a metallic film. By controlling the chain/surface distance we may load, store and release photons in the atoms, making them a promising tool for future quantum memories.
\end{abstract}
\maketitle

\section{Introduction}
Advances in ordered atomic arrays \cite{Chang2018,Rui2020,Ruostekoski2023} have recently highlighted the potential of subwavelength-spaced quantum emitters as a versatile platform for engineering cooperative optical phenomena \cite{Asenjo2017,Bettles2016a}. The cooperative nature of subradiant states, arising from destructive interference among the emitters' radiation fields, leads to markedly reduced decay rates and long coherence times, enabling applications ranging from efficient photon storage \cite{Porras2008,Ferioli2021} to quantum sensing \cite{Ostermann2013,Facchinetti2016,Zafra2026}. Yet the very mechanism that protects these states from radiative decay also makes their direct excitation and selective preparation intrinsically challenging. In free space, existing strategies for accessing subradiant states include the use of spatially tailored driving fields that match the symmetry of dark collective modes \cite{Asenjo2017,Olmos2013}, local phase and amplitude imprinting \cite{Jenkins2012,Facchinetti2016}, and adiabatic protocols that transfer population from bright to subradiant manifolds \cite{Ruostekoski2017,Asenjo2017}. These approaches typically rely on engineering the spatial phase profile of the collective excitation or on exploiting coherent couplings between bright and dark modes, making the preparation of the most subradiant states increasingly demanding as the system size grows. 

Within finite one-dimensional chains of two-level atoms with spacing $a<\lambda_0/2$, where $\lambda_0=2\pi/k_0$ is the resonant transition wavelength, subradiant states in the $k$-space of the first Brillouin zone lie outside the light cone $k=k_0$ \cite{Asenjo2017,Needham2019,Piovella2024}. Because these states reside beyond the light line ($k>k_0$), their direct preparation in a two-level atomic chain cannot be achieved using conventional optical excitation. A strategy could be to drive a localized section of the array, which will excite many modes at the same time \cite{Piovella2025b}: at later time, only the subradiant modes will survive with a substantial population. However, very few photons are emitted through this protocol (subradiant modes get weakly populated only), making this configuration challenging experimentally. Alternatively, atomic chains coupled to one-dimensional waveguides have been proposed as a promising alternative \cite{Ruostekoski2017,Asenjo2017,Masson2020}, where guided photons naturally couple to collective excitations with wave vectors beyond the light line, enabling the preparation and manipulation of long-lived subradiant and selectively radiant states through waveguide-mediated interactions.

Here we propose an efficient way to address selectively subradiant modes with $k>k_0$ by setting the array close to a metallic substrate and exploit the evanescent field generated in the total internal reflection or in surface plasmonic (SP) wave. A recent work \cite{Araujo2018} has shown that cooperative effects, expected in free space, are absent when the atoms are close to a surface whose polariton resonances coincide with the dominant atomic dipole coupling. In this case, cooperative effects are overshadowed by the very fast decay of the atomic fluorescence into surfaces modes.
However, we show that this is not the case when the atoms interact not with polariton resonances but with a SP wave propagating along the interface vacuum-metal. Whereas in free space light can be emitted only radiatively, surface provides the possibility of non-radiative emission into bound surface waves. Their normal component of the wavevector is purely immaginary, whereas the parallel component has a value larger than the free space, $k>k_0$. The atoms are driven by the evanescent wave with a wave vector matching the subradiant mode with $k>k_0$ and a single collective subradiant state is populated in which the excitation remains for long times. The surface plasmon is excited by a laser incident in total internal reflection condition from a dielectric substrate that carries the thin metallic film (see Fig.\ref{fig1}). The coupling between atoms and SP provides an extra decay channel other that the free space one. Excitation by the plasmon electric field and the decay to the free and surface modes lead to a stationary state where the subradiant mode is populated, with a linewidth depending on the finite size of the chain and on the metal losses.
In the following, we introduce the model describing atomic chains coupled to a metallic surface carrying the propagating SP. Then, we study the dynamics of the atoms using the coupled-dipole model in the single-excitation limit and the generation of subradiant modes.

\section{Model}

\subsection{Atom arrays in free space}

We consider $j=1,\dots,N$ two-level atoms with transition frequency $\omega_0=ck_0$ between the ground state $|g_j\rangle$ and the excited state $|e_j\rangle$, transition dipole moments $\mathbf{d}_j=d\mathbf{\hat d}_j$ and positions $\mathbf{r}_j$.
In regimes where the Born-Markov approximation works well, one can trace out the quantized light fields and obtain the field-mediated dipole-dipole
couplings between the emitters described by an effective Hamiltonian \cite{Lehmberg1970,Akkermans2008}
\begin{equation}
H=-\hbar\sum_{j,m=1}^N \mathbf{\hat d}^*_j\cdot \mathbf{G}(\mathbf{r}_j-\mathbf{r}_m)\cdot \mathbf{\hat d}_m\sigma^\dagger_j\sigma_m,\label{Heff}
\end{equation}
where $\sigma_j=|g_j\rangle\langle e_j|$ and $\mathbf{G}$ is the dyadic Green's tensor in the 3D free space, given by
\begin{equation}
\mathbf{G}(\mathbf{r})=\frac{3\Gamma_0}{2}\frac{e^{ik_0r}}{k_0r}\left\{\mathbf{I}-\mathbf{\hat r}\mathbf{\hat r}+\left(\mathbf{I}-3\mathbf{\hat r}\mathbf{\hat r}\right)\left(\frac{i}{k_0r}-\frac{1}{k_0^2r^2}\right)\label{G0}
\right\}
\end{equation}
where $\mathbf{I}$ is the identity tensor $3\times 3$,  $\mathbf{\hat r}=\mathbf{r}/r$ with $r=|\mathbf{r}|$ and $\Gamma_0=\omega_0^3d^2/3\pi\hbar\epsilon_0 c^3$ is the single-atom spontaneous decay rate. We consider atom arrays along the $x$-axis in 3D free space, where atoms are equally separated by $a$ and transition dipoles are all oriented in the same direction $\mathbf{\hat d}$. Restricting our analysis to the case of a single excitation shared among the atoms, the Hamiltonian describes transitions among the states $|j\rangle=|g_1,\dots,e_j,\dots,g_N\rangle$. For an infinite array, $N\rightarrow \infty$, the Hamiltonian (\ref{Heff}) has
singly excited eigenstates in the form of Bloch states, $|k\rangle_\infty=\sum_j e^{ika(j-1)}|j\rangle$ with $k\in [-\pi/a,\pi/a]$ and complex eigenvalues $\Omega_k-i\Gamma_k$. Notably, $\Gamma_k$ vanishes outside the interval $[-k_0,k_0]$ \cite{Asenjo2017,Cech2023}. For a finite array, we may expand the singly excited eigenstates of $H$ on these states restricted to the $N$ lattice sites and normalized. We refer to these states as the 'quasi-Bloch' states, defined as \cite{Zhang2020}
 \begin{equation}
 |k\rangle =\frac{1}{\sqrt{N}}\sum_{j=1}^N e^{ika(j-1)}|j\rangle\label{Dicke}
 \end{equation}
Although the states $|k\rangle$ are neither eigenstates of $H$ neither hortogonal, nevertheless they form an over-complete basis in the single-excitation manifold. We introduce the cooperative decay rate spectrum as $\Gamma(k)=-(1/\hbar)\mathrm{Im}\langle k|H|k\rangle$ \cite{Piovella2024}, mimicking the exact spectrum of eigenvalues of $H$ for large values of $N$.
In finite arrays, states (\ref{Dicke}) with $k\notin[-k_0,k_0]$ are identified as subradiant states with tiny but finite decay rates.

\subsection{Effect of the surface}

We now consider an atom chain in vacuum parallel and close to the plane surface of a metallic film with dielectric constant $\epsilon_1(\omega)$ and thickness $d$, on top of a substrate with dielectric constant $\epsilon_2$ (see Fig.\ref{fig1}). The atom chain is at a distance $h$
from the surface at $z=0$ of the metal, occupying together with the dielectric substrate the lower half-space $z\le 0$. The upper half-space $z>0$ is the vacuum, with $\epsilon_0=1$. A resonant $p$-polarized laser beam, with frequency $\omega=\omega_0$, is incident  from the bottom on the interface between the dielectric and the metal with an angle $\theta$ larger than the limit angle. Such a beam is able to excite a SP wave propagating along the metal/vacuum interface with wave vector $k_x=k_0\sqrt{\epsilon_1/(1+\epsilon_1)}$. If $\epsilon_1$ is negative then $k_x$ is larger than the vacuum value $k_0$ and the SP wave number matches a subradiant mode around $k=k_x$ in the atomic chain, with an evanescent wave amplitude decreasing exponentially in the upper half-space $z>0$ as $\exp(-\xi z)$ with $\xi=\sqrt{k_x^2-k_0^2}$. Generally, $\epsilon_1$ has an imaginary part due to finite conductibility in the metal and the SP wave is attenuated when is propagating along the interface at $z=0$. The detailed expressions for the transmission coefficient of the laser through the two interfaces and the SP wave are reported in the Supplemental Material (SM) \cite{SM}.
\begin{figure}
      \centerline{\scalebox{0.9}{\includegraphics{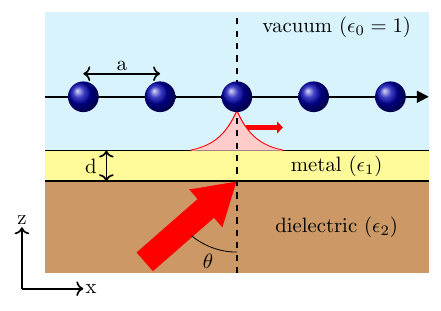}}}
        \caption{\textbf{Setup:} The setup consists of a three-layered structure in Kretschmann configuration and an atom chain placed in the vacuum near the metallic surface. At a specific angle, an incidence laser on the dielectric can resonantly excite a SP wave propagating in the $x-$direction and confined at the interface between the metal and the vacuum.}\label{fig1}
\end{figure}
The presence of the surface changes the mode structure of the electromagnetic field, having a number of effects on the atomic interaction and emission \cite{Jones2018}. For atom-surface distances on the order of the transition wavelength and greater, the dominant effect is interference between the emitted and reflected fields. Constructive and destructive interference leads to an increase or decrease in the spontaneous decay rate, respectively. As the atom-surface distance is reduced, nonradiative processes become more significant, leading to an increase in the spontaneous decay rate for very small distances.
The presence of a surface can be accounted for by adding to the free-space Green’s tensor $\mathbf{G}(\mathbf{r}_j-\mathbf{r}_m)$ of Eq.(\ref{G0}) a scattering term of the form \cite{Sipe1981,Jones2018}
\begin{equation}
\mathbf{G}_S(\mathbf{r}_j-\mathbf{r}_m)=\frac{i\Gamma_0}{4\pi}\int_0^\infty dk_\rho\frac{k_\rho}{k_z}\left(
\mathbf{G}^s-\frac{k_z^2}{k_0^2}\mathbf{G}^p\right)\label{Gs}
\end{equation}
where $\mathbf{G}^s$ and $\mathbf{G}^p$ represent the contributions of the $s$-wave and $p$-wave, respectively, $k_z=\sqrt{k_0^2-k_\rho^2}$ and $k_\rho=\sqrt{k_x^2+k_y^2}$ is the component of the wave vector parallel to the surface. The expressions of $\mathbf{G}^s$ and $\mathbf{G}^p$ are reported in the SM \cite{SM}.
Fig.\ref{fig2} shows $\Gamma(k) / \Gamma_0$ vs. $ka\in[-\pi,\pi]$ for a chain of $N=20$ atoms in free-space (blue line) and at a distance $h=\lambda_0/\pi$ from the surface (yellow line) for dipoles parallel (left) and perpendicular (right) to the surface. The horizontal dashed black line is the single-atom decay rate in free-space ($\Gamma=\Gamma_0$), while the two dashed red lines are the states with $k=\pm k_0$. The presence of the surface tends to partially cancel the collective behavior of the atoms.
\begin{figure}
      \centerline{\scalebox{0.3}{\includegraphics{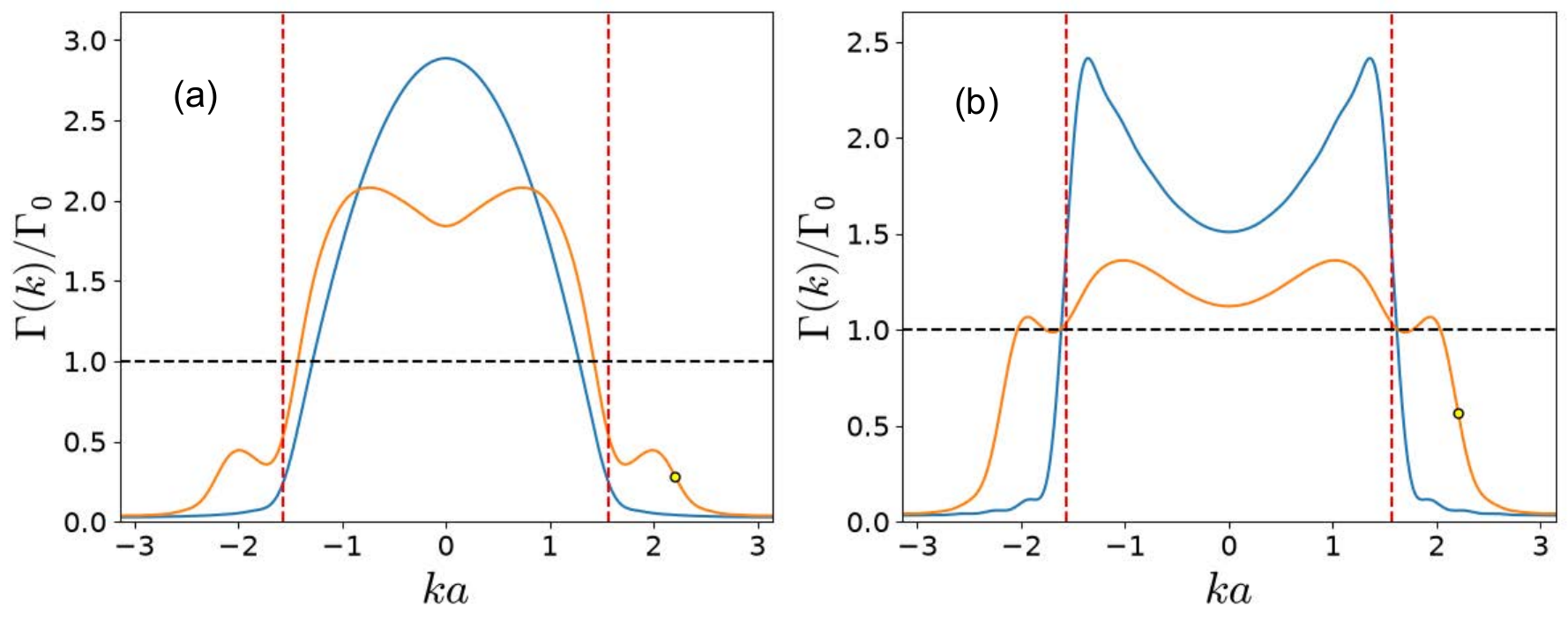}}}
        \caption{$\Gamma(k) / \Gamma_0$ vs. $ka$ for a chain of $N=20$ atoms in free-space (blue line) and at a distance $h=\lambda_0/\pi$ from the surface (yellow line). (a): dipoles parallel to the surface.  (b): dipoles orthogonal to the surface. The horizontal dashed black line corresponds to the single-atom decay rate in free-space, while the two vertical dashed red lines are the states with $k=\pm k_0$.}\label{fig2}
\end{figure}

\subsection{Dynamics}

The dynamics of the excited state of the $N$ atoms in the chain, driven by the SP wave and interacting with the vacuum and surface modes, are described by the coupled-dipole model described by the following equations \cite{Araujo2018},
 \begin{equation}
 \dot\beta_j=\left(i\Delta-\frac{\Gamma}{2}\right)\beta_j-i\frac{\Omega_{SP}(z)}{2}e^{ia(j-1)}-\frac{1}{2}\sum_{m=1\atop m\neq j}^N G_{j,m}\beta_m
 \label{beta_dot:vec}
  \end{equation}
with $\beta_j=\langle\sigma_j\rangle$. This model assumes that most atoms are in the ground state ($\langle\sigma^z_j\rangle\approx -1$) and neglects correlations ($\langle\sigma^z_j\sigma_j\rangle\approx \langle\sigma^z_j\rangle\langle\sigma_j\rangle$). The coupling term 
$G_{j,m}=\mathbf{\hat d}\cdot \mathbf{G}(\mathbf{r}_j-\mathbf{r}_m)\cdot \mathbf{\hat d}=G_{j,m}^{(0)}+G_{j,m}^{(S)}$ includes the interaction with the vacuum modes, $G_{j,m}^{(0)}$ given by Eq.(\ref{G0}), and the interaction with the surface, $G_{j,m}^{(S)}$ given by Eq.(\ref{Gs}).
The detuning is $\Delta=\Delta_0+\delta$ with $\Delta_0=\omega-\omega_0$, where $\omega$ is the frequency of the driven field and $\delta=\mathrm{Re}\left[G_{j,j}^{(S)}\right]$ is the Casimir-Polder shift \cite{Araujo2018} (the Lamb shift $\mathrm{Re}\left[G_{j,j}^{(0)}\right]$ is embodied in the bare resonant frequency $\omega_0$). The single atom decay is $\Gamma=\Gamma_0+\Gamma_S$ where $\Gamma_S=\mathrm{Im}\left[G_{j,j}^{(S)}\right]$  is the surface-induced decay rate. Finally, $\Omega_{SP}(z)=t_{210}\Omega_0e^{-\xi z}$, with $\Omega_0=\mathbf{d}\cdot \mathbf{E}_{0}/\hbar$ the Rabi frequency associated to the electric field amplitude of the SP wave propagating along the interface vacuum/metal ($x$-axis). This wave is generated by a monochromatic laser with electric field amplitude $ \mathbf{E}_{0}$, incident from the medium with dielectric constant $\epsilon_2$ with the resonant surface-plasmon angle $\theta_{sp}$, $p$-polarized in the plane $(x,z)$ and with transmission coefficient $t_{210}$ through the two interfaces.
We assume the atomic state in the single-excitation manifold, $|\Psi\rangle=\alpha|G\rangle+\sum_{j=1}^N\beta_j|j\rangle$ (with $|G\rangle=|g_1,\dots,g_N\rangle$ the ground state, $\alpha\approx 1$ and $\beta_j\ll 1$). The mean excitation probability is $E = (1/N)\sum_{j=1}^N|\beta_j|^2$. By defining the projection of the atomic state on the quasi-Bloch state $|k\rangle$ as $A_N(k)=\langle k|\Psi\rangle=(1/\sqrt{N})\sum_j\beta_j e^{-ika(j-1)}$, the normalized probability distribution is $P(k)=|A_N(k)|^2/\int_{-\pi/a}^{\pi/a}|A_N(k)|^2dk$.

\section{Results}
We consider a chain of $N=100$ atoms in vacuum, modeled as dipoles with atomic transition wavelength at $\lambda_0=780 \mathrm{nm}$, and we investigate both parallel and transverse dipole orientations. The SP wave is generated using a multilayer structure composed of a metallic film of thickess $d=100\ \mathrm{nm}$ and a dielectric layer with permittivities $\epsilon_1=-2+0.1\, i$ and $\epsilon_2 = 3.0625$, respectively. The atom chain is at a distance $h=\lambda_0/\pi$ from the the metal surface and driven by the SP field generated by a weak ($\Omega_0 = 0.1\Gamma_0$) and far-detuned laser ($\Delta = 10 \Gamma_0$), incident on the dielectric at the plasmon resonance angle (see Fig.\ref{fig1}). As shown in Fig.\ref{fig3}, the evolution of the mean excitation probability $E$ reaches a steady-state value after a sufficient time for both dipole orientations. These values are orders of magnitude smaller than those obtained under direct laser excitation (with the same parameters $\Omega_0=0.1\Gamma_0$ and $\Delta = 10\Gamma_0$), reflecting the exponential decay of the SP field amplitude with the atom-surface distance.
To assess the subradiant character of the prepared steady state, we report the free evolution of $E$ after the laser has been switched off and the surface moved away (Fig.\ref{fig4}). For a better comparison, the values of $E$ are divided by the steady-state values $E_{ss}$ at $\Gamma_0t=50$. The evolution does not show the characteristic superradiant decay, typical for laser driving in free-space and remain almost constant over long times, suggesting that the state excited through SP is composed almost entirely of subradiant modes. This is particularly evident for transversely oriented dipoles, which couple more efficiently to the evanescent wave due to the polarization nature of the SP electric field \cite{Sipe1981}. 
The distribution of excitation in momentum space (see Fig.\ref{fig5}) provides information about the mode population and shows that we selectively excite a 'quasi-Bloch' state in the subradiant region of the spectrum around $k=k_x$. The asymmetric and broadened character of the distribution is due to both finite size effects and metal losses (see SM \cite{SM}). If the surface is now kept near the chain when the laser is turned off (see Fig.\ref{fig6}), its effect on the system is both to partially suppress the cooperative behavior of the emission and to enhance the single-atom decay rate. As the distance decreases, this results in a progressive loss of the subradiant character of the excitation. SPs therefore provide a mechanism to preserve the collective subradiant state, while the atom-surface separation acts as a extra decay channel enabling the stored excitation to be released on demand. 
\begin{figure}
      \centerline{\scalebox{0.45}{\includegraphics{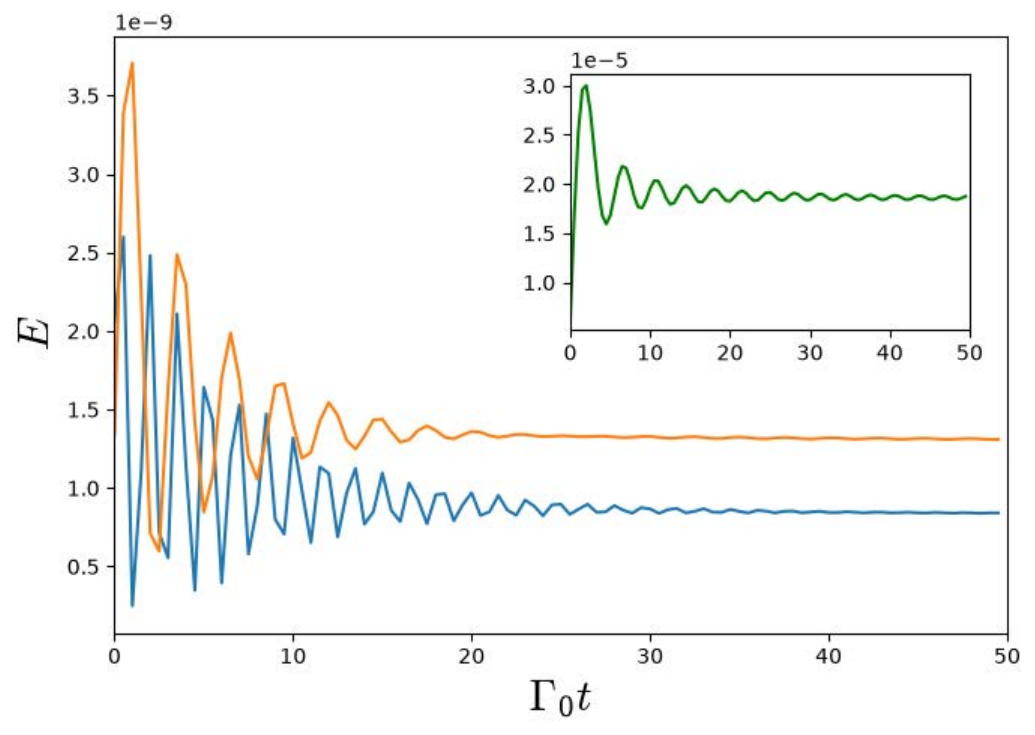}}}
        \caption{Evolution of the mean excitation probability $E$ under SP driving for $N=100$ atoms for parallel (blue line) and perpendicular (yellow line) dipole orientations to the surface. The SP wave is generated by a weak ($\Omega_0=dE_0/\hbar =0.1\Gamma_0$) and far-detuned ($\Delta=10\Gamma_0$) laser. The metal has $\epsilon_1=-2+0.1\,i$ and thickness $d=100\ \mathrm{nm}$, and the atom chain is at a distance $h=\lambda_0/\pi$ from the surface.  The inset shows the results for the chain driven in free space by a laser with the same $\Omega_0$.}\label{fig3}
\end{figure}
\begin{figure}
     \centerline{\scalebox{0.45}{\includegraphics{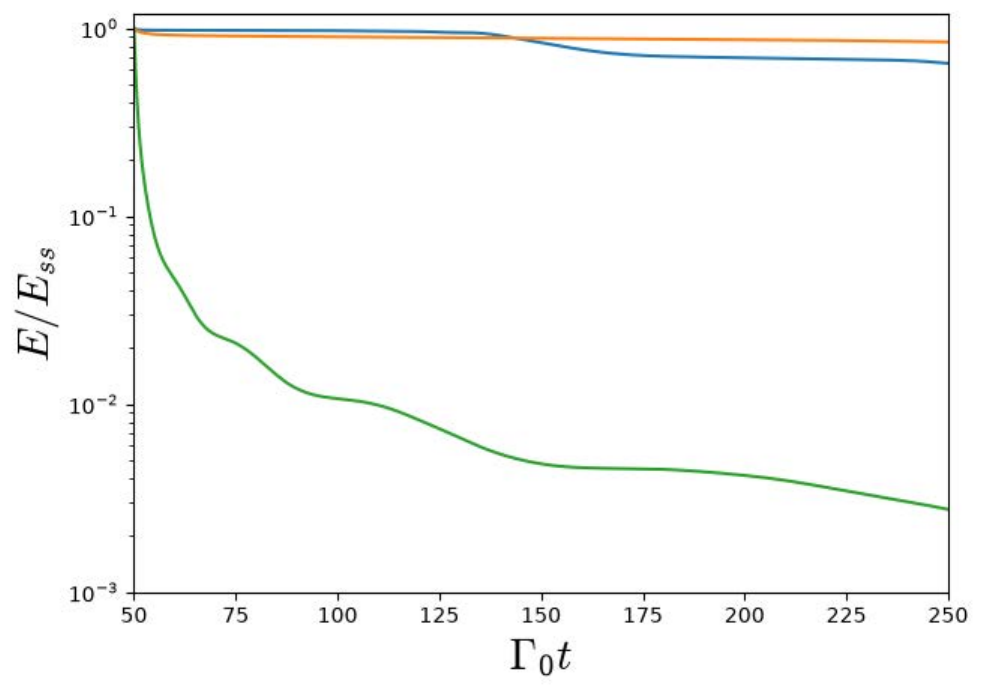}}}
        \caption{Free evolution of the normalized mean excitation probability $E/E_{ss}$ for the three cases of Fig.\ref{fig3}, after that the laser is switched off at $\Gamma_0t=50$.
%a chain of $N=100$ atoms is driven up to steady state ($\Gamma_0 t= 50$), where the laser is switched off and the surface moved away and let it evolves freely. \textbf{Green line:} Chain in free-space driven by a laser. \textbf{Blue and Yellow line:} Chains driven by SPs with dipoles oriented parallel and transverse to the surface respectively.
         }\label{fig4}
\end{figure}
\begin{figure}
\centerline{\scalebox{0.35}{\includegraphics{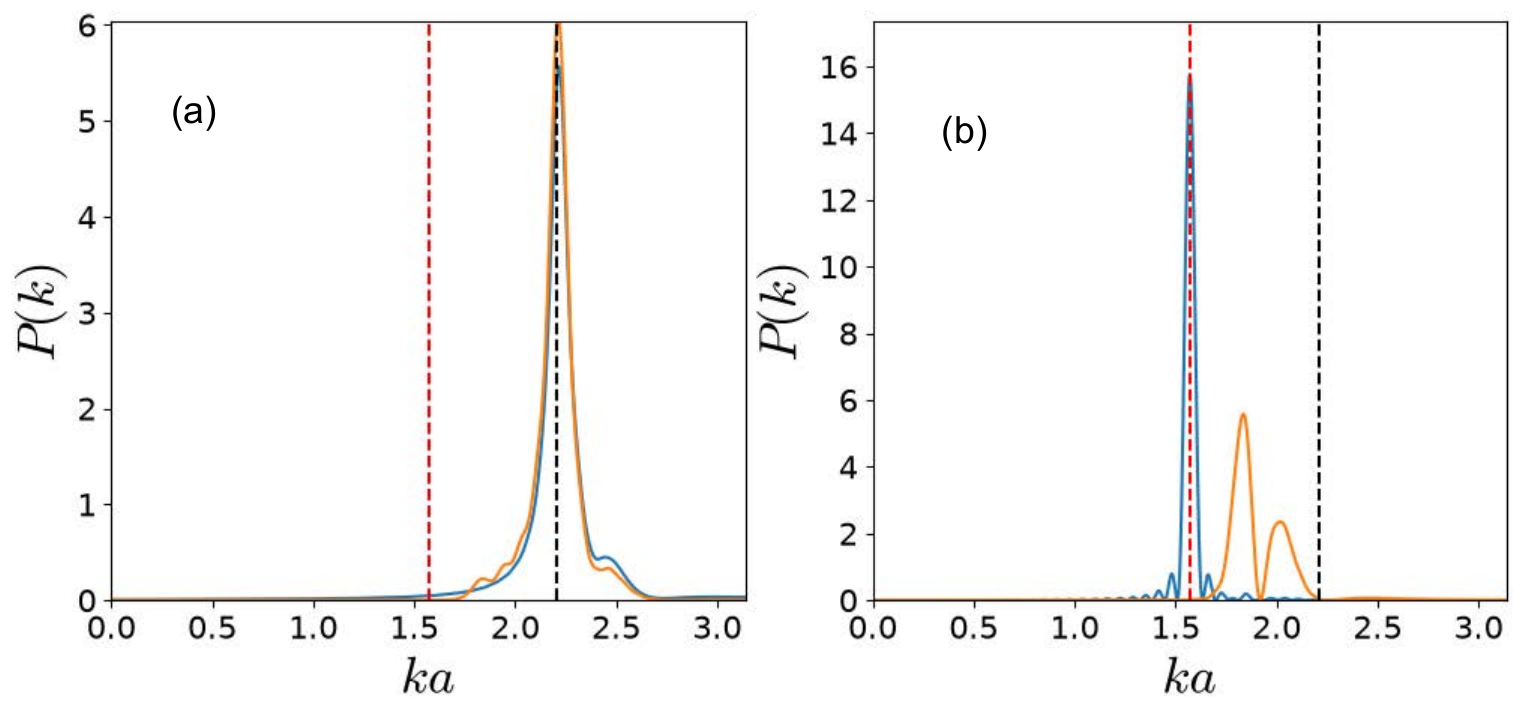}}}
        \caption{Distribution of the excitation in the momentum space at $\Gamma_0 t= 50$ (blue line) and at $\Gamma_0t = 100$ (yellow line) for SP driving (a) and for direct laser driving (b). The red dashed line represents the mode with $k=k_0$ and in black the one with $k=k_x$.}\label{fig5}
\end{figure}
\begin{figure}
      \centerline{\scalebox{0.45}{\includegraphics{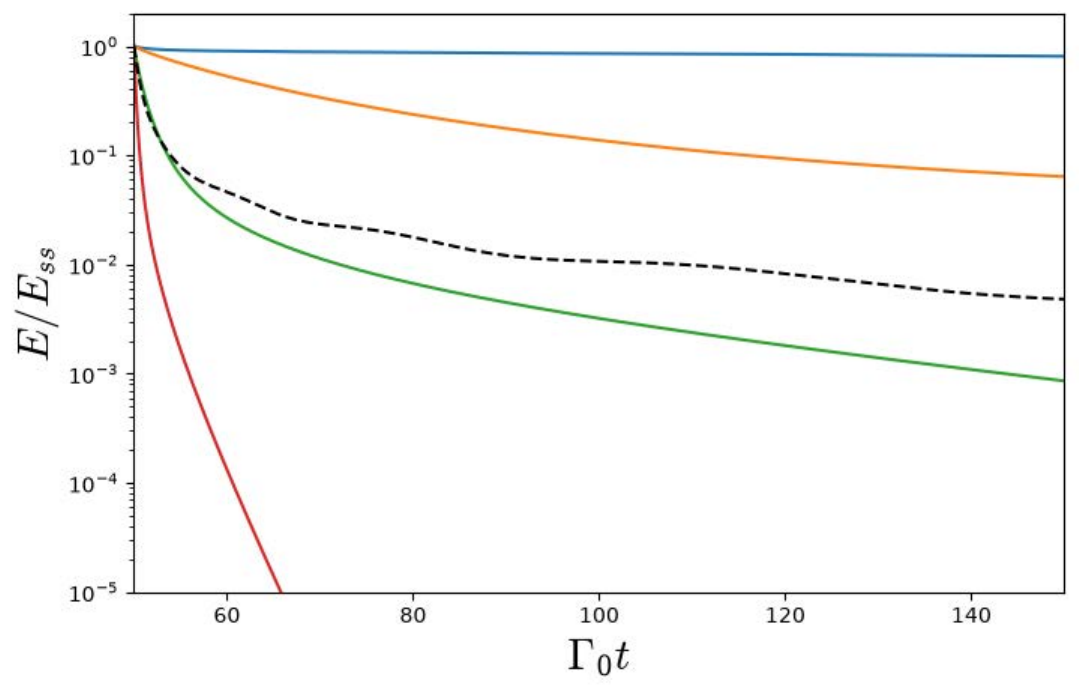}}}
        \caption{Evolution of the normalized mean excitation probability $E/E_{ss}$ for different atom-surface distances. After the steady state is reached (at $\Gamma_0t=50$), the laser is turned off and the surface kept fixed in its position. From top to bottom: $h = \lambda_0, \frac{\lambda_0}2, \frac{\lambda_0}{4}, \frac{\lambda_0}{8}$. The black dashed curve represents the laser driving case in free-space. The other parameters are the same as in Fig.\ref{fig3}.}\label{fig6}
\end{figure}
 
\section{Conclusions}

In conclusion, we have shown how directly addressing subradiant states by SP waves. By supplying the momentum lacking in free-space optical excitation, SPs enable direct coupling to highly subradiant collective modes of ordered atomic arrays. Furthermore, the atom-surface distance provides a simple experimental knob for tuning the coupling and selectively preparing long-lived collective excitations. These findings offer a realistic strategy for accessing and exploiting subradiant states, transforming them from intrinsically dark modes into experimentally controllable resources for collective quantum optics and quantum technologies.
From an experimental perspective, two aspects are particularly relevant. First, since most dielectric materials do not support surface-plasmon resonances at optical or near-infrared frequencies, nanophotonic metasurfaces may be employed to engineer resonances at the desired atomic transition frequency \cite{Aljunid2016,Laliotis2021}. Second, the atom-surface distance must be controlled with sufficient precision, as it governs the coupling to the plasmonic mode. This can be achieved through tunable trapping potentials, for example by applying external electric fields. In addition, switching off the driving laser or rotating its polarization from $p$ to $s$ suppresses plasmon generation, providing a simple mechanism to interrupt the excitation process and rapidly transfer population out of the surface channel. Together, these capabilities make the proposed approach experimentally accessible with state-of-the-art platform technologies.

%\acknowledgements
%The authors wish to thank P.W. Courteille for helpful discussions.

\newpage

\begin{widetext}

\section{Supplementary Material for: \\ Excitation of subradiant states in atom chains by surface plasmon waves}

\subsection{Reflection and Transmission coefficients} \label{App:A}

We consider a three-layer system composed by a plane interface (along the $x$-axis) between vacuum (with $\epsilon_0=1$) and a metallic layer of thickness $d$ with dielectric constant $\epsilon_1$  on top of a dielectric substrate with dielectric constant $\epsilon_2$. The reflection and transmission coefficients of a $p$-polarized electromagnetic wave incident on the dielectric with $\epsilon_2$ are:
\begin{eqnarray}   
    r_{210}&=&\frac{r_{21}+r_{10}e^{2ik_{1z}d}}{1+r_{21}r_{10}e^{2ik_{1z}d}}\\
    t_{210}&=&\frac{t_{21}t_{10}e^{ik_{1z}d}}{1+r_{21}r_{10}e^{2ik_{1z}d}}
\end{eqnarray}
where $r_{21}=(\epsilon_1k_{2z}-\epsilon_2k_{1z})/(\epsilon_1k_{2z}+\epsilon_2k_{1z})$, $t_{21}=2\epsilon_2k_{2z}/(\epsilon_1k_{2z}+\epsilon_2k_{1z})$, $r_{10}=(\epsilon_0k_{1z}-\epsilon_1k_{0z})/(\epsilon_0k_{1z}+\epsilon_1k_{0z})$, $t_{10}=2\epsilon_1k_{0z}/(\epsilon_0k_{1z}+\epsilon_1k_{0z})$ and $k_{iz}=(\omega\sqrt{\epsilon_i}/c)\cos\theta_i$ with $i=0,1,2$ and $\theta_i$ are the angles to the normal at the two interfaces. If the laser is incident at the interface between $2$ and $1$ with a wave vector $\mathbf{k}_2=(k_{2x},0,k_{2z})$, then $k_{1z}=\sqrt{(\omega/c)^2\epsilon_1-k_{2x}^2}$ and $k_{0z}=\sqrt{(\omega/c)^2-k_{2x}^2}$.

If $\epsilon_1$ is real, the surface-plasmon resonance angle corresponds to the value for which the reflection coefficient $r_{10}$ becomes infinite i.e. when 
$\epsilon_0k_{1z}+\epsilon_1k_{0z}=0,$ from which it follows
\begin{equation}
k_{0x}=k_{1x}=k_{2x}=\frac{\omega}{c}\sqrt{\frac{\epsilon_1}{1+\epsilon_1}}\label{k_SP}
\end{equation}
and $k_{0z}=\sqrt{(\omega/c)^2-k_{0x}^2}=(\omega/c)\sqrt{1/(1+\epsilon_1)}$. Hence, if $\epsilon_1<-1$ then $k_{0z}=i\xi$ is imaginary and the surface plasmon wave is confined to the interface between the vacuum and the metal, where it propagates with a wave vector $k_{0x}>(\omega/c)$.
Adopting the Lorentz-Drude model for the metal, its dielectric constant is $\epsilon_1(\omega)=1-\omega_p^2/(\omega^2+i\gamma\omega)$ where 
$\omega_p=\sqrt{e^2 n_e/m\epsilon_0}$ is the plasma frequency, $n_e$ is the electron density and $\gamma$ is the loss parameter, related to the conductivity $\sigma$ by the relation $\sigma=\epsilon_0\omega_p^2/(\gamma-i\omega)$. Writing $\epsilon_1=\epsilon_1'+i\epsilon_1''$, we have
\begin{equation}
\epsilon_1'=1-\frac{\omega_p^2}{\omega^2+\gamma^2}\quad,\quad \epsilon_1''=\frac{\gamma\omega_p}{\omega^2+\gamma^2}
\end{equation}
and the inverse expressions
\begin{equation}
\omega=\omega_p\frac{\sqrt{1-\epsilon_1'-\epsilon_1''}}{|1-\epsilon_1'|}\quad,\quad \gamma=\omega_p\frac{\epsilon_1''}{1-\epsilon_1'}.
\end{equation}
The surface plasmon wavenumber (\ref{k_SP}) as a functon of the frequency $\omega$ is
\begin{equation}
k_x(\omega)=\frac{\omega}{c}\sqrt{\frac{\omega^2-\omega_p^2+i\gamma\omega}{2\omega^2-\omega_p^2+2i\gamma\omega}}.
\end{equation}
Fig.\ref{fig1} shows the real and imaginary part of $k_x(\omega)/k_0$ (where $k_0=\omega/c$) vs $\omega/\omega_p$. The peak of $\epsilon_1'$ is near to $\omega=(\omega_p/\sqrt{2})(1-\epsilon_1''/2)^{1/2}$, which corresponds to the polariton with $\epsilon_1'=-1$.
\begin{figure}
      \centerline{\scalebox{0.45}{\includegraphics{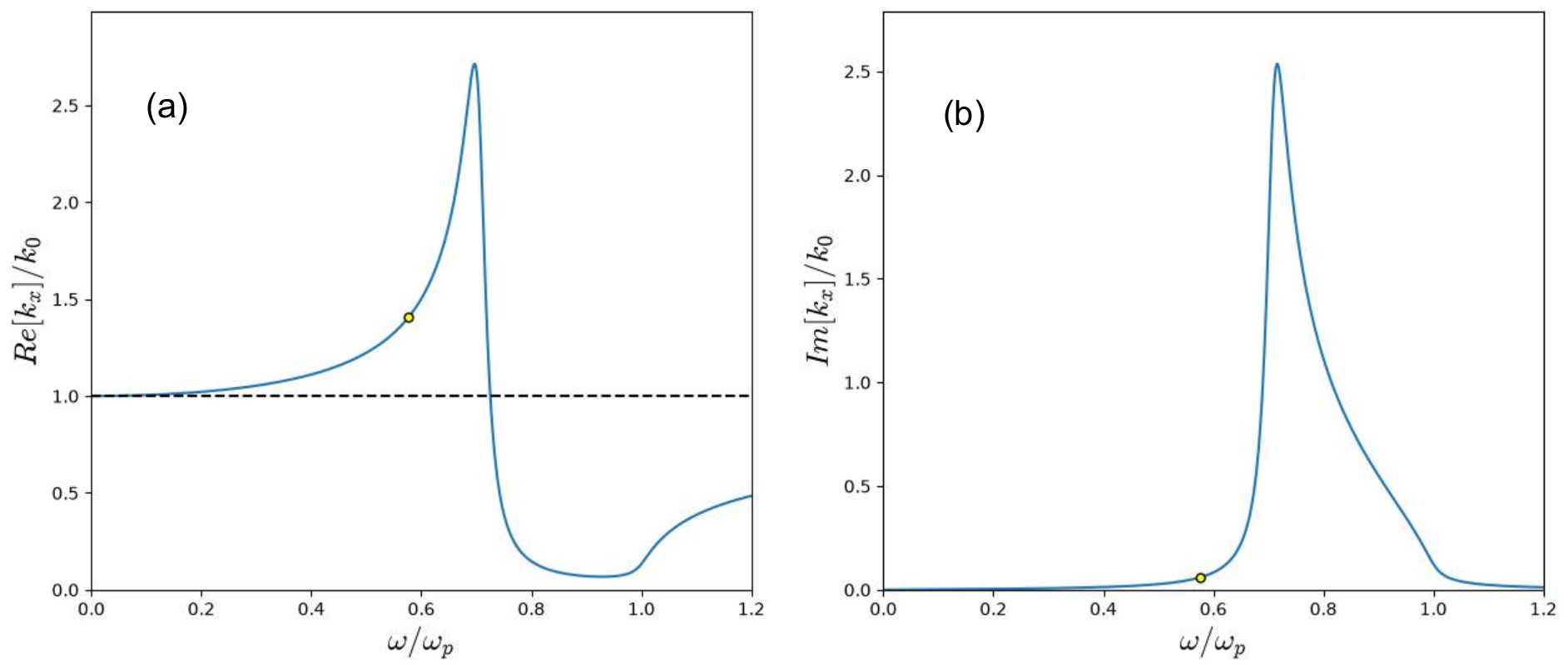}}}
     \caption{Real (a) and Imaginary part (b) of $k_x(\omega)$ vs $\omega/\omega_p$, with $\gamma=0.033\omega_p$. The yellow dots in both figures represent the values for the frequency considered for the surface plasmon in the main paper ($\omega = 0.576\omega_p$), corresponding to $\epsilon_1=-2+i\,0.1$.}\label{fig1}
\end{figure}

\subsection{Scattering Green Tensor}\label{App:B}
In cartesian coordinates, the components of the two tensors $\textbf{G}^s$ and $\textbf{G}^p$ are given by \cite{Sipe1981,Jones2018}
\begin{equation}
\begin{aligned}
    \textbf{G}^s_{xx}=\frac{A^s}{2}[J_0(k_{\rho}\rho) + J_2(k_{\rho}\rho) \cos(2\phi)], \hspace{0.5cm} \textbf{G}^s_{yy} = \frac{A^s}{2}[J_0(k_{\rho}\rho) - J_2(k_{\rho}\rho) \cos(2\phi)], \hspace{0.5cm} \textbf{G}^s_{zz}=0,\\
    \textbf{G}^s_{xy}=\textbf{G}^s_{yx}=-\frac{A^s}{2}J_2(k_{\rho}\rho)\sin(2\phi), \hspace{0.5cm}  \textbf{G}^s_{xz}=\textbf{G}^s_{zx}=0, \hspace{0.5cm} \textbf{G}^s_{yz}=\textbf{G}^s_{zy} = 0
\end{aligned}
\end{equation}
    \begin{equation}
\begin{aligned}
    \textbf{G}^p_{xx} = \frac{A^p}{2}[J_0(k_{\rho}\rho) - J_2(k_{\rho}\rho) \cos(2\phi)], \hspace{0.5cm} \textbf{G}^p_{yy} = \frac{A^p}{2}[J_0(k_{\rho}\rho) + J_2(k_{\rho}\rho) \cos(2\phi)], \hspace{0.5cm} \textbf{G}^p_{zz} = -\frac{k_{\rho}^2}{k_z^2}A^pJ_0(k_{\rho}\rho), \\
    \textbf{G}^p_{xy}=\textbf{G}^p_{yx}=\frac{A^p}{2}J_2(k_{\rho}\rho)\sin(2\phi), \hspace{0.5cm} \textbf{G}^p_{xz}=-\textbf{G}^p_{zx}=i\frac{k_{\rho}}{k_z}A^pJ_1(k_{\rho}\rho)\cos(\phi), \hspace{0.5cm} \textbf{G}^p_{yz}=-\textbf{G}^p_{zy}=i\frac{k_{\rho}}{k_z}A^pJ_1(k_{\rho}\rho)\sin(\phi)
\end{aligned}  
\end{equation}     
where $\rho = \sqrt{x_{jm}^2+y_{jm}^2}$ is the distance between the atoms in the plane parallel to the surface, $\phi = \arccos(\frac{x_{jm}}{\rho})$ and $J_n$ are the Bessel functions of order $n$. We also define:
\begin{equation}
\begin{aligned}
    A^s = r_{012}^se^{2ik_{0z}h} \hspace{0.3cm},  \hspace{0.3cm} A^p = r_{012}^pe^{2ik_{0z}h}
\end{aligned}
\end{equation}
with $h$ the atom-surface distance. The functions $r_{012}^s$, $r_{012}^p$ are the Fresnel reflection coefficients of the three-layered structure for $s$- and $p$-polarization, respectively, defined as: 
\begin{equation}
\begin{aligned}
    r_{012}^{\alpha} = \frac{r_{01}^{\alpha} + r_{12}^{\alpha}e^{2ik_{1z}d}}{1 + r_{01}^{\alpha}r_{12}^{\alpha}e^{2ik_{1z}d}} \\
\end{aligned}
\end{equation}
with ${\alpha} = \{s, p\}$, $d$ the thickness of the metal film and
\begin{equation}
\begin{aligned}
    r_{01}^s=\frac{k_z - k_{1z}}{k_{0z} + k_{1z}}, \hspace{0.5cm} r_{12}^s=\frac{k_{1z} - k_{2z}}{k_{1z} + k_{2z}} \\
\end{aligned}
\end{equation}
\begin{equation}
\begin{aligned}
     r_{01}^p=\frac{\epsilon_1k_{0z} - k_{1z}}{\epsilon_1k_{0z} + k_{1z}}, 
     \hspace{0.5cm} r_{12}^p=\frac{\epsilon_2k_{1z} - \epsilon_1k_{2z}}{\epsilon_2k_{1z} + \epsilon_1k_{2z}}
\end{aligned}
\end{equation}
To numerically evaluate the integral in Eq.(4) we used an elliptical integration path in the complex plane of $k_{\rho}$ to avoid the singularities in $k_{0z}$ and in the reflection coefficients, as done by Paulus et al. \cite{Paulus2000}.

\subsection{Losses and Finite size effects}\label{App:C}
The collective atomic steady state excited by surface plasmons is not just the single 'quasi-Bloch' mode with $k=k_x$, but it is characterized by a peak with finite width, as the distribution of the excitation probability in momentum space shows in Fig.5 of the main paper. This results both from losses in the metal, responsible for the decay of the plasmonic field in the $x$-direction, and from size effects due to the finite number of atoms composing the chain. Fig.\ref{fig2} shows the FWHM $\Delta k$ vs $N$, (a), and vs $Im[\epsilon_1]$, (b). By fitting the values obtained, we find that the width of the peak can be expressed as a function of $N$ and $Im[\epsilon_1]$ as:
\begin{equation}
\begin{aligned}
     \Delta k = \frac{a}{N} + b\{Im[\epsilon_1]\}^c
     \label{fit}
\end{aligned}
\end{equation}
with $a=5.5$, $b=1.7$ and $c=1.75$. These parameters are almost independent on the metal thickness $d$ and on the distance $h$ of the chain from the metal surface, as shown in Fig.\ref{fig3}.
\begin{figure}
     \centerline{\scalebox{0.5}{\includegraphics{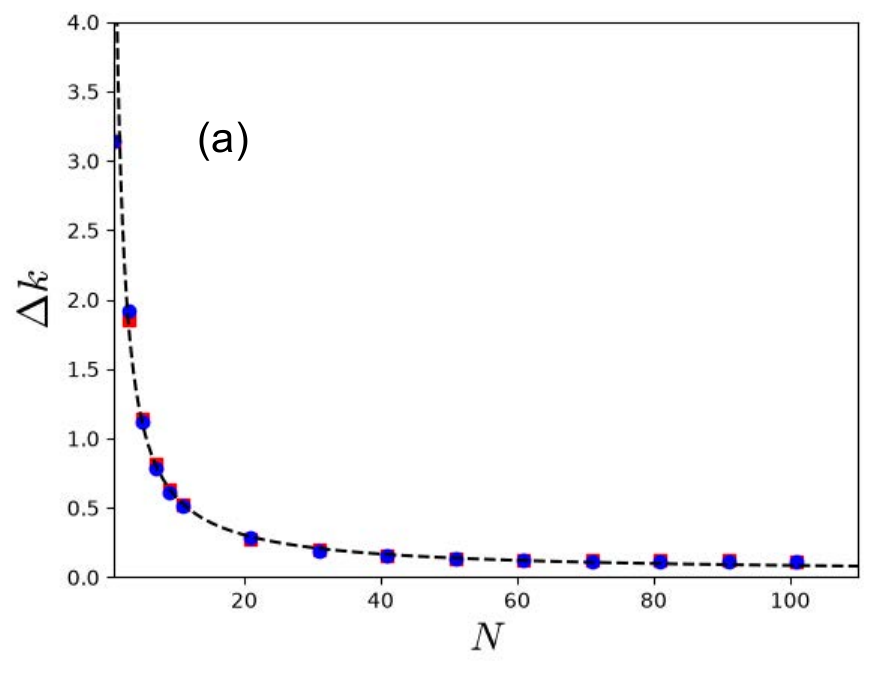}\includegraphics{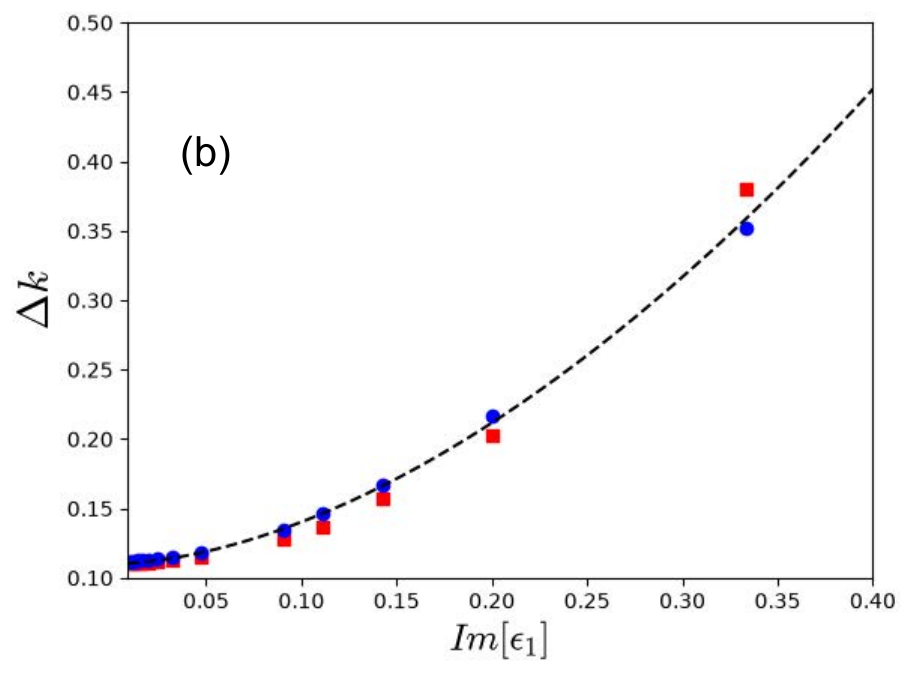}}}
        \caption{(a): FWHW of the excitation probability distribution in momentum space $\Delta k$ at steady state ($\Gamma_0t=50$) vs the number $N$ of atoms for both parallel (red squares) and perpendicular dipoles (blue dots) and $\epsilon_1=-2+0.1 i$. (b) FWHM vs  the imaginary part of the metal permittivity $Im[\epsilon_1]$, with the same other parameters as in (a): $d = 100\ \mathrm{nm}$ and $h = \lambda_0/\pi$. The black dashed line is the fit of Eq.(\ref{fit}).}
        \label{fig2}
\end{figure}
\begin{figure}
     \centerline{\scalebox{0.5}{\includegraphics{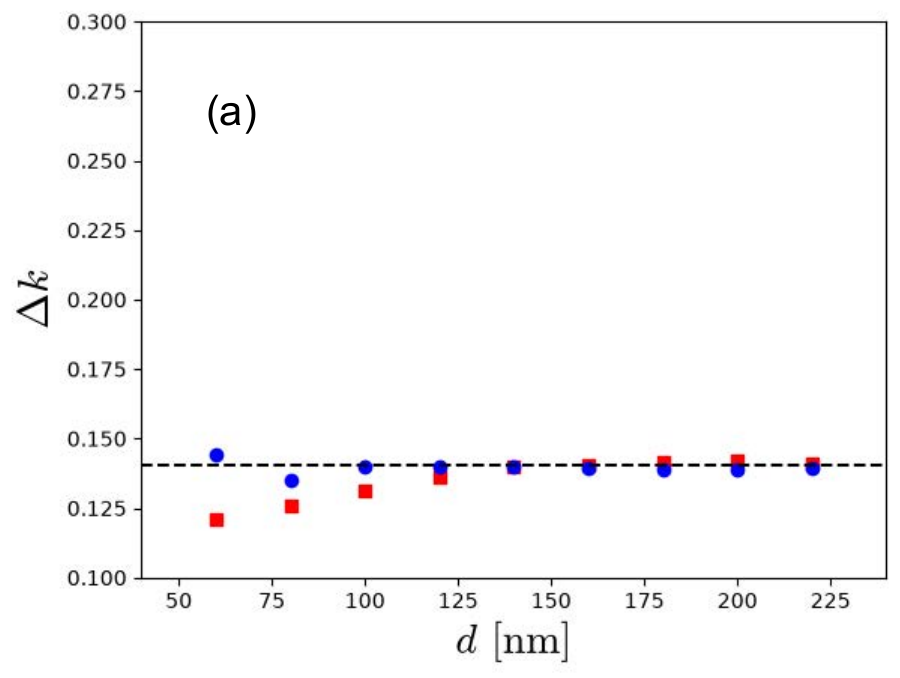}\includegraphics{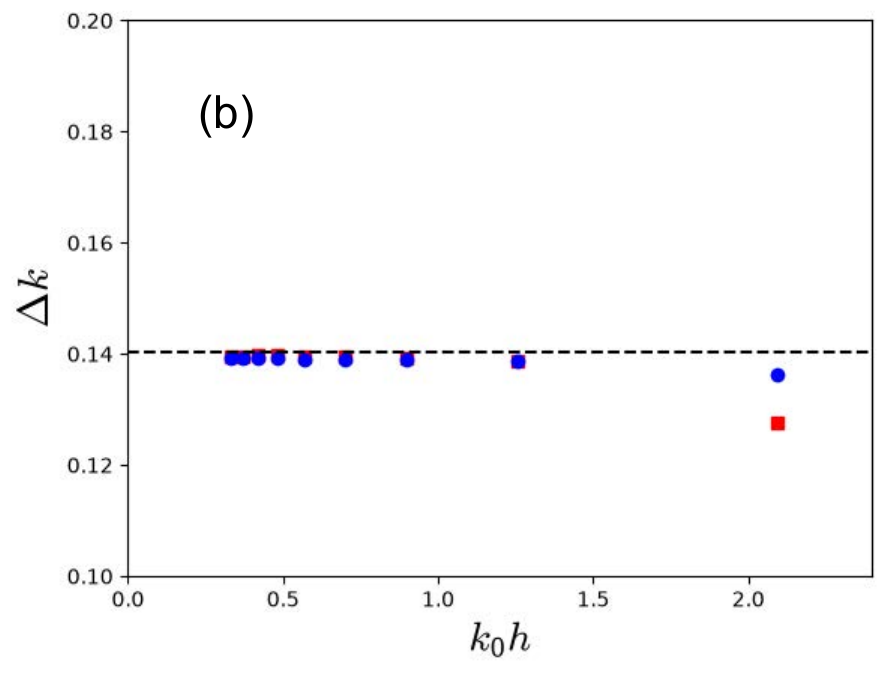}}}
        \caption{(a): FWHW $\Delta k$ of the excitation probability distribution in momentum space  at steady state ($\Gamma_0t=50$) (a) vs the metal thickness $d$ and (b) vs the distance $h$ of the chain from the metal surface, for $N=50$, parallel (red squares) and perpendicular dipoles (blue dots) and for $\epsilon_1=-2+0.1 i$. The black dashed lines represent the value of Eq.(\ref{fit}) for this case.}
        \label{fig3}
\end{figure}

\end{widetext}


\begin{thebibliography}{999}
\bibitem{Chang2018} D. E. Chang, J. S. Douglas, A. González-Tudela, et al. Colloquium: Quantum Matter Built from Nanoscopic Lattices of Atoms and Photons, Rev. Mod. Phys. \textbf{90}, 031002 (2018).
\bibitem{Rui2020} J. Rui, D. Wei, A. Rubio-Abadal, S. Hollerith, J. Zeiher, D.M. Stamper-Kurn, C. Gross, I. Bloch. A Subradiant Optical Mirror Formed by a Single Structured Atomic Layer,  Nature \textbf{583}, 369 (2020).
\bibitem{Ruostekoski2023} J. Ruostekoski. Cooperative Quantum-Optical Planar Arrays of Atoms, Phys. Rev. A \textbf{108}, 030101 (2023).
\bibitem{Asenjo2017} A. Asenjo-Garcia, M. Moreno-Cardoner, A. Albrecht, 
H.J. Kimble, D.E. Exponential improvement in photon storage fidelities using subradiance and “selective radiance” in
atomic arrays, Phys. Rev. X \textbf{7}, 031024 (2017).
\bibitem{Bettles2016a} R.J. Bettles, S.A. Gardiner, C.S. Adams. Cooperative eigenmodes and scattering in one-dimensional atomic arrays, Phys. Rev. A {\bf 94}, 043844 (2016).
\bibitem{Porras2008} D. Porras, J.I. Cirac. Collective generation of quantum states of light by entangled atoms, Phys. Rev. A {\bf 78}, 053816 (2008).
\bibitem{Ferioli2021} G. Ferioli, A. Glicenstein, L. Henriet, I. Ferrier-Barbut, and A. Browaeys, Storage and release of subradiant excitations
in a dense atomic cloud, Phys. Rev. X {\bf 11}, 021031 (2021).
\bibitem{Ostermann2013} L. Ostermann, H. Ritsch, and C. Genes. Protected state enhanced quantum metrology with interacting two-level ensembles. Phys. Rev. Lett. \textbf{111}, 123601 (2013).
\bibitem{Facchinetti2016} G. Facchinetti, S.D. Jenkins, J. Ruostekoski. Storing light with subradiant correlations in arrays of atoms, Phys.Rev. Lett. {\bf 117}, 243601 (2016).
\bibitem{Zafra2026} Diego Zafra-Bono, Oriol Rubies-Bigorda and S.F. Yelin. Subradiant collective states for precision sensing via transmission spectra, Phys. Rev. A {\bf 114}, 013704 ( 2026).
\bibitem{Olmos2013} B. Olmos, D. Yu, Y. Singh, P. Schmelcher and M. Saffman. Long-range interacting many-body systems with alkaline-earth-metal atoms, Phys. Rev. A {\bf 87}, 023604 (2013).
\bibitem[12]{Jenkins2012} S.D. Jenkins, J. Ruostekoski. Controlled manipulation of light by cooperative response of atoms in an optical lattice, 
Phys. Rev. A {\bf 86}, 031602(R) (2012).
\bibitem{Ruostekoski2017} J. Ruostekoski and J. Javanainen.  Proposes transferring excitations into subradiant collective modes of atoms coupled to a 1D waveguide, effectively storing light in a dark state, Phys. Rev. A \textbf{96}, 033857 (2017).
\bibitem{Needham2019} J.A. Needham, I. Lesanovsky, B. Olmos. Subradiance-protected excitation transport, New J. Phys. {\bf 21}, 073061 (2019).
\bibitem{Piovella2024} N. Piovella. Cooperative Decay of an Ensemble of Atoms in a One-Dimensional Chain with a Single Excitation,  Atoms {\bf 12}, 43 (2024).
\bibitem{Piovella2025b} N. Piovella and R. Bachelard. Subradiant Decay in 2D and 3D Atomic Arrays, Photonics {\bf 12}, 1214 (2025).
\bibitem{Masson2020} S. J. Masson and A. Asenjo-Garcia. Interprets ordered atomic arrays themselves as atomic waveguides supporting non-radiative guided modes, Phys. Rev. Research 2, 043213 (2020). 
\bibitem{Araujo2018} M.O. Ara\`{u}jo, J. C. de Aquino Carvalho, Ph.W. Courteille and A. Laliotis. Cooperative atomic emission from a line of atoms interacting with a resonant plane surface, Phys. Rev. A {\bf 110}, 032813 (2024).
\bibitem{Lehmberg1970} R.H. Lehmberg. Radiation from an $N$-atom system. I. General formalism, Phys. Rev A \textbf{2}, 883 (1970).
\bibitem{Akkermans2008} E. Akkermans, A. Gero, R. Kaiser. Photon localization and Dicke superradiance in atomic gases, Phys. Rev. Lett. {\bf 101},  103602 (2008).
\bibitem{Cech2023} M. Cech, I. Lesanovsky, B. Olmos. Dispersionless subradiant photon storage in one-dimensional emitter chains, Phys. Rev. A {\bf 108}, L051702 (2023).
\bibitem{Zhang2020} Y.X. Zhang and K. M\"{o}lmer. Subradiant Emission from Regular Atomic Arrays: Universal Scaling
of Decay Rates from the Generalized Bloch Theorem, Phys. Rev. Lett. {\bf 125}, 253601 (2020).
\bibitem{SM} See Supplemental Material at [URL will be inserted by publisher] for  multilayer transmission coefficient, surface plasmon wave, surface Green function and width of the spectrum distribution.
\bibitem{Jones2018} R. Jones, J.A. Needham, I. Lesanovsky, F- Intravaia, and B. Olmos. Modified dipole-dipole interaction and dissipation in an atomic ensemble near surfaces, Phys. Rev. A {\bf 97}, 053841 (2018).
\bibitem{Sipe1981} J.E. Sipe, the dipole antenna problem in surface physics: a new approach, Surface Science \textbf{105}, 489 (1981).
\bibitem{Aljunid2016} S.A. Aljunid, E.A. Chan, G. Adamo, M. Ducloy, D. Wilkowski and N.I. Zheludev, Nano Lett. \textbf{16}, 3137 (2016).
\bibitem{Laliotis2021} A. Laliotis, B.S. Lu, M. Ducloy and D. Wilkowski, Atom-surface physics: A review. AVS Quantum Sci. \textbf{3}, 043501 (2021).
\end{thebibliography}

\begin{thebibliography}{999}
\bibitem{Sipe1981} J.E. Sipe, the dipole antenna problem in surface physics: a new approach, Surface Science \textbf{105}, 489 (1981).
\bibitem{Jones2018} R. Jones, J.A. Needham, I. Lesanovsky, F- Intravaia, and B. Olmos. Modified dipole-dipole interaction and dissipation in an atomic ensemble near surfaces, Phys. Rev. A {\bf 97}, 053841 (2018).
\bibitem{Paulus2000} M. Paulus; P. Gay-Balmaz; O.J.F. Martin. Accurate and efficient computation of the Green's tensor for stratified media. {\em Phys. Rev. E} {\bf 2000}, {\em 62}, 5797-5807.
\end{thebibliography}
\end{document}